\documentclass[12pt, preprint2]{aastex}
\usepackage{epsfig}

\begin{document}

\title{An unified  model for superluminal motion and state transition
 on  microquasars and quasars}

\author{B.P.\ Gong\altaffilmark{1,2}}
\altaffiltext{1}{Centre d'Etude Spatiale des Rayonnements,
CNRS-UPS, 9, avenue du Colonel Roche, 31028 Toulouse Cedex 4,
France} \altaffiltext{2}{Department of Physics, Huazhong
University of Science and Technology, Wuhan 430074, P.R. China
}

\begin{abstract}
Superluminal motion has been interpreted as relativistically
moving out flow. The increasing observations of  such sources
provide chance to investigate their mechanism in detail. This
paper proposes that superluminal motion may be jet precession
induced blob motion. This model can interpret a number of
observational phenomena which are not well understood under the
scenario of bulk motion, such as the absence of the receding blob
after the outburst
 and the receding blob becomes brighter than the
approaching one in  XTE J1550-564 .
Moreover the compton scattering of photos from a warped
 accretion disk may form a structured jet. And the precession of
the structured jet may lead to the complicated state transitions
observed in microquasars like  GRS 1915+105. Applying to  AGNs,
the model may explain  the scales of length and time of the
phenomena proportioning  to the mass of the black hole. The new
model can be tested easily on microquasars, i.e.,  XTE J1550-564,
on which  the previous receding blob should move towards the core
instead of away from it after 2002, and the previous approaching
blob shall disappear.



\end{abstract}




\section{Introduction}

The appearance on the sky of relativistically moving out flow, which
expands at speed greater than the speed of light has been
predicted by Rees (1966) five years in advance of the discovery
of superluminal motion~\citep{whitney71,  cohen71}.

The apparent velocity of
separation is the ratio of the difference in observed positions to the
observed time-interval.
Suppose that at $t=0$ in our frame, the moving blob is at the original
point, a distance $D$ away from us, and sends us a signal. That
signal arrives at the time $D/c$. After time $\Delta t$ in our frame,
the moving blob has moved a distance $v\Delta t \sin\theta$ in the
transverse direction and sends us a second signal. This one arrives
at $t=\Delta t +D/c-(v/c)\Delta t \cos \theta$. Thus the apparent
transverse velocity is,
\begin{equation}
\label{vapp2}
v_{a}=\frac{v\sin\theta}{1-\frac{v}{c}\cos\theta}
 \,,
\end{equation}
when $v\sim c $ and $\theta \sim 0$, $v_{a}\sim 2c/\theta $, which
can be as large as 10 times the speed of light, as shown in  Fig.~\ref{Fig:f1}a.

The first superluminal source discovered in our galaxy was
 GRS 1915+105 a recurrent transient source of hard X-ray found by
Castro-Tirado et al 1994. Soon after similar
phenomenon was observed by two different groups GRO
J1655-40~\citep{zhang94}, XTE J1748-288~\citep{smith98}.
These galactic source are called  microquasars
in  which the Black Hole (BH) is only a few solar masses instead of
several million solar masses as in the case of quasars.

Mirabel et al (1994)
observed with the very Large Array (VLA) four ejection events of
radio-emitting clouds from the high energy source GRS 1915+105. Who
also  noted small
apparent changes in the projection of the jets on the
sky~\citep{mira99}.

Fender et al. (1999)
reported observations of a major outburst with the Multi-Element
radio Linked Interferometer Network  (MERLIN), which is five times
the angular resolution of the VLA at 5 GHz. These observations
revealed again relativistic ejections, but with
  higher (by approximately $20\%$) proper motion than
those measured with VLA.

Miller-Jones et al. (2005)
reported two separate outbursts in 2001 March and 2001 July. As
the former observations, no deceleration is observed in  GRS
1915+105. Liner polarization is observed in the approaching jet
component, with a gradual rotation in position angle and a
decreasing fractional polarization with time. These observations
are difficult to understand under the scenario of  bulk motion.

 On the other hand,  there are striking
similarities between GRS 1915+105 and the SS 433~\citep{mira99},
in  which the jet precession scenario has been widely accepted.

\begin{figure}[t]
\includegraphics[87, 217][600, 600]{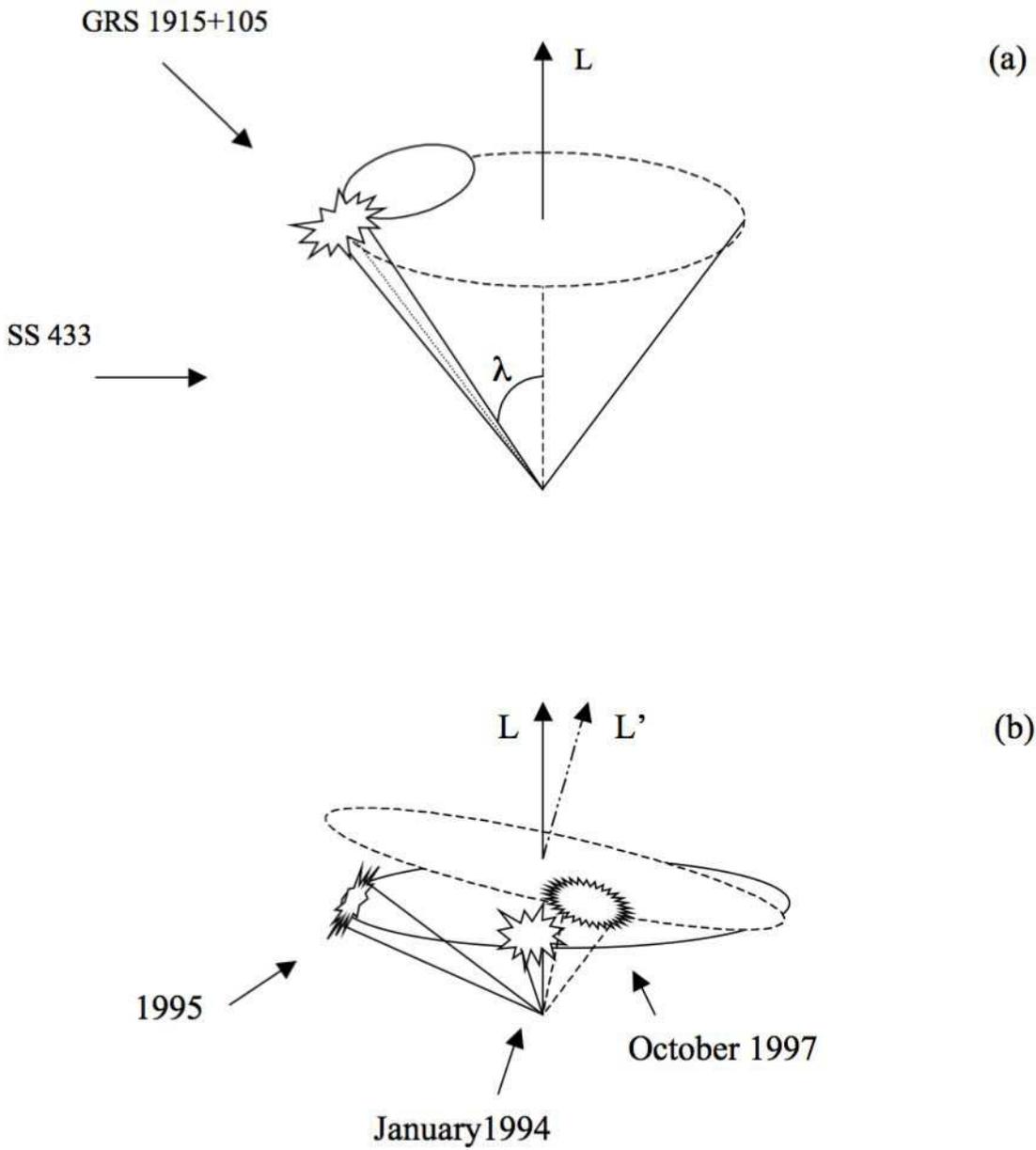}
\caption{\label{Fig:f2} A schematic illustrating of the jet precession model. (a)
shows that GRS 1915+105 different from SS433 only in the view angle,
the former has chance to see the emission corresponding to  $\theta\sim 0$,
whereas the latter never does.
The motion of the jet is  a primary precession with a opening angle
of  precession cone,  $\lambda$ (doted ellipse),  superimposed by a
nutation with small opening angle (solid ellipse).
(b) The outburst of  January 1994 corresponds to the starting of
period of precession. And the outburst of October  1997 corresponds
the beginning of another period. The position angle may vary by both
short-term effect,  $\sim P_{b}$, and long-term effect,  $\sim 3.8$ yr. }
\end{figure}

This paper suggests that   superluminal motion on microquasars
like GRS 1915+105  may  also be an effect of jet precession,
differing  from  SS 433  only on the misalignment angle between
the jet axis and Line of Sight (LOS). In the precession of SS 433,
the angle between the jet axis and  LOS is relatively larger,
whereas, in GRS 1915+105, the jet axis precesses across the LOS,
as shown in Fig.~\ref{Fig:f2}a. The jet precession model can
interpret following phenomena:


(a) why the ratio of the
displacement of the approaching blob to the  receding blob is not
a constant.

(b) why the receding jet appears  later than  the approaching jet
in  XTE J1550-564 (for year).

(c) why the position angle changes $\sim 10$deg in 20-30-d in GRS
1915+105.

(d) why there is obvious oscillation in the receding blob.

(e) why $20\%$ discrepancy in proper motion between the ejection of
1994 and 1997.

(f) why deceleration appeared in  XTE J1550-564  but not in   GRS
1915+105.

(g) why the decrease in flux density with angular separation from
the core on  GRS 1915+105  is remarkably similar with that of SS
433.

Thanks to the rapid jet precession in microquasars, the new model can
be tested easily, i.e.,
 the return of the approaching of GRS 1915+105 should be  observed
 in around one year after each  major
 outburst.

 Microquasar  XTE J1550-564  may already reverse its  approaching and
receding blob through precession, hence
the previously receding blob should move towards the core after 2002.
Correspondingly the Doppler-shift feature of  XTE J1550-564  in year 2002 should has
opposite sign comparing  with that of year 2000.

 Jet precession induced blob motion  may explain
the  superluminous motion. Moreover, the   precession and nutation
of a structured jet
can interpret  the state transition and oscillation observed in
microquasars like  GRS 1915+105.



\section{The model}


The jet-disk interaction is still an open question. In this paper
we assume that the precession of the accretion disk causes the
change of the jet axis. Consequently the jet axis varies  relative
to LOS.

Similar as SS433, the jet precesses in certain half-opening angle
cone, the individual jet components move ballistically away from
the system~\citep{mig05}. The direction in which the blobs are
observed to moving match the direction of the jets at the time
when the blobs were just forming.
And the travel direction of the blobs may  indicate the average
phase angle of the jet during the time of their
formation~\citep{dunn06}. Therefore, the length of the blobs to
the core is approximately a constant  by assuming an approximately
equal ISM density around a microquasar.

With $D_{0}$ the distance of the observer to the approaching blob
(when the position angle $\theta=0$),  the instantaneous distances
of the approaching and receding blobs relative to the observer are
given
\begin{equation}
\label{dadr} D_{a}=D_{0}+l_{0}\sin\lambda[1-\cos\theta_{a}], \ \
D_{r}=D_{0}+l_{0}\sin\lambda[1+\cos\theta_{r}] \,,
\end{equation}
where $\theta_{a}\equiv\theta(T_{a})$
  is  the misalignment angle between the axis of the approaching jet
and  LOS,  and $\theta_{r}\equiv\theta(T_{r})$ corresponds to  that of the receding jet
(note $\theta_{r}$ and $\theta_{a}$ denote absolute values).
 In which $T_{a}$ and  $T_{r}$ are the time of arrivals of the two blobs
in observer's frame,  $l_{0}$ is the distance from the core to
each blob, and $\lambda$  is the misalignment angle between the
jet axis and the orbital angular momentum vector, ${\bf L}$, as
shown in Fig.~\ref{Fig:f1}b.

The blobs appeared to move away from the core. The separation of the
approaching and receding blobs measured by the observer are given by,
\begin{equation}
\label{flux} R_{a}=l_{0}\sin\lambda\sin\theta_{a}  \,, \ \ \
R_{r}=l_{0}\sin\lambda\sin\theta_{r}  \,,
\end{equation}
The angles,  $\theta_{a}$ and $\theta_{r}$ can be denoted by
$\theta$, which varies with  the precession of the jet
axis~\citep{dunn06},
\begin{equation}
\label{chobs} \cos \theta=\cos i\cos\lambda+\sin i \sin\lambda \cos\eta \,,
\end{equation}
where  $i$ is the orbital inclination angle.  And
$\eta=\dot{\Omega}_{1} t+\eta_{0}$ is the precession phase of the
jet axis, in which  $\dot{\Omega}_{1}$ is the precession frequency
of accretion disk by~\citep{katz97},
\begin{equation}
\label{katz}
\dot{\Omega}_{1}=-\frac{3}{4}\frac{Gm_{2}}{a}
(\frac{a_{d}}{a})^2\frac{\cos\lambda}{(Gm_{1}a_{d})^{1/2}}
 \,,
\end{equation}
where $m_{1}$ and  $m_{2}$ are the masses of the accreting object and
the companion star respectively, $a$ and $a_{d}$ are the separation
of the two bodies and the disk radius respectively.

\begin{figure}[h]
\includegraphics[47, 37][500, 500]{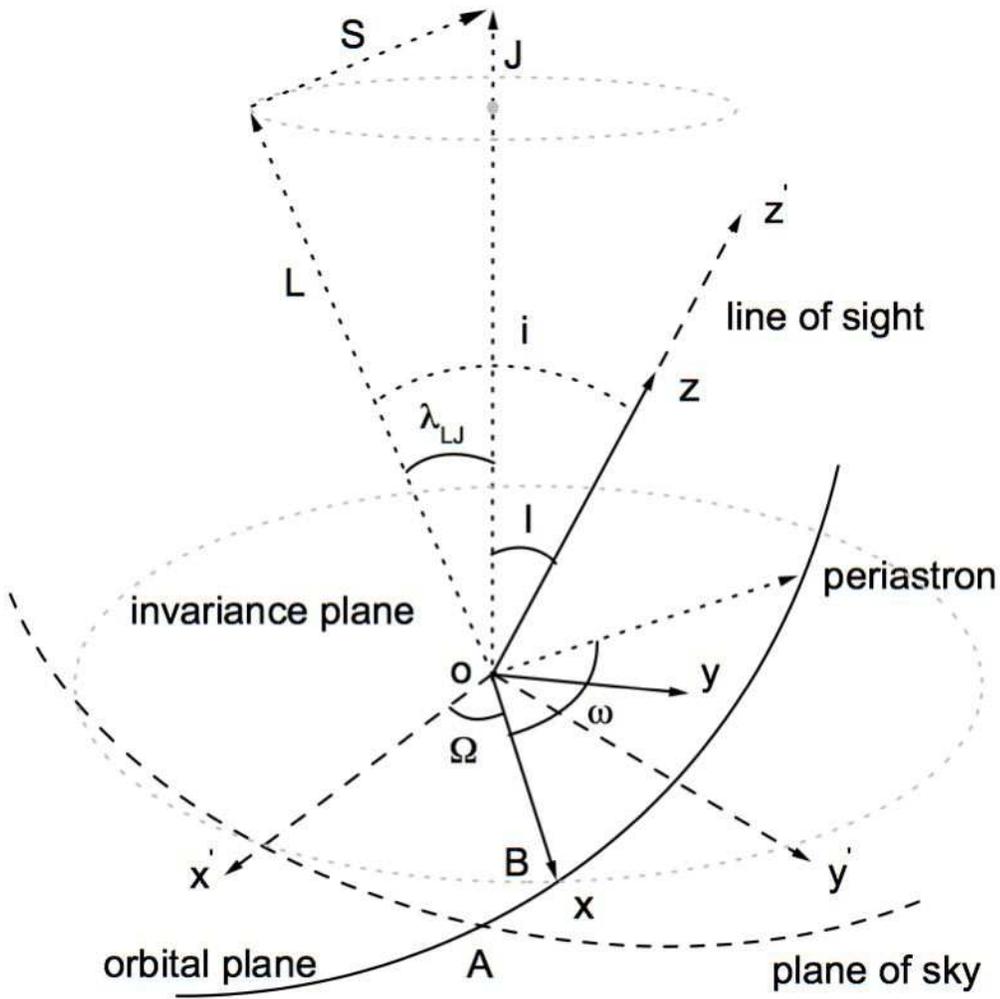}
\caption{\label{Fig:f3} Definition of angles and reference frames.
The ascending node,  $o-x$,  rotates with
respect to the fixed line,  $o-x^{\prime}$. The coordinate system,
$x$,  $y$,  $z$ rotates with respect to the static coordinate
system,  $x^{\prime}$,  $y^{\prime}$,  $z^{\prime}$. }
\end{figure}

Actually Eq.~($\ref{katz}$) is equivalent to a general expression,
Eq.~(47) of Barker and
O'Connell (1975),
which    can be understands as
 a result of perturbation by the mass distribution of the disk
 to the classical gravitational two-body
 problem. The perturbation  also affects the six orbital
 elements, $i$, $e$, $a$, $\Omega$, $\omega$, $P_{b}$. In which $e$
 is the eccentricity of the orbit, $P_{b}$ is the orbital period,
 $\omega$ is the longitude of the periastron, and $\Omega $
 is  the longitude of the ascending node, as shown in  Fig.~\ref{Fig:f3}.

 The precession of the disk results the precession of the
orbital plane, therefore, the ascending node is not at
static  with respect to the LOS,
thus  $i$ in Eq.~($\ref{chobs}$) is a function of time,
\begin{equation}
\label{i} \cos i =\cos I\cos\lambda_{LJ}+\sin I \sin\lambda_{LJ}
\cos \Omega  \,,
\end{equation}
where $I$ is the misalignment
angle between the
total angular momentum and LOS,  and $\lambda_{LJ}$ is the misalignment
angle  between ${\bf L}$ and  ${\bf J}$, as shown in Fig.~\ref{Fig:f3}.

 The general expression of $\dot{\Omega}$ is given by
Barker $\&$ O'Connell (1975),
the magnitude of which along the total angular momentum is
given~\citep{wk},
 \begin{equation}
\label{bo}
\dot{\Omega}=\frac{3}{4}\frac{GmJ_{2}}{a^{5}(1-e^{2})^{2}}
\frac{P_{b}}{2\pi }\frac{\sin 2\lambda}{\sin \lambda_{LJ}}
 \,,
\end{equation}
where $J_{2}=(I_{3}-I_{1})/m_{d}$ ($I_{3}$ is the principal moment
of inertia of the disk,  $I_{1}$ is the moment of inertia with
respect to an arbitrary axis, and $m_{d}$ is the disk mass).

Notice that $\dot{\Omega}$ can be  larger than $\dot{\Omega}_{1}$.
Thus the variation of  $\theta$ depends not only on the precession
phase, $\eta$ of Eq.~($\ref{chobs}$), but also on the  rapid
variation of the orbital inclination angle, $i$, induced by
$\dot{\Omega}$ of Eq.~($\ref{bo}$).  The precession of $\eta$
denotes the precession of jet along the precession cone with
opening angle $\lambda$ at time scale, i.e., years, and the
variation of $i$ leads to a nutation with much smaller opening
angle, and at shorter time scale, as shown by the small solid
ellipse in Fig.~\ref{Fig:f2}a.

\section{Application to  super luminous motion}
\label{sec:slm}

By the given orbital parameters  $P_{b}=33.5\pm1.5$-d,
$m_{1}=14\pm 4M_{\odot}$, $m_{2}=1-1.5M_{\odot}$~\citep{fender04},
 we have $a= 8\times 10^{12}$cm. And assuming  $a_{d}=a/2$ and
$\lambda=\pi/6$, obtains,  $\dot{\Omega}_{1}\sim 0.26$deg
day$^{-1}$ by Eq.~($\ref{katz}$). In such  case the period of
$\eta$ is approximately $\sim 2\pi/\dot{\Omega}_{1}\approx$3.8 yr.

If the  outbursts   correspond
to $\theta\sim 0$ (the jet axis is approximately aligned with LOS),
as shown in  Fig.~\ref{Fig:f2}b,
then the 3.8yr period of  the jet precession implies that the
outbursts of  GRS 1915+105 in January 1994~\citep{mira94}, October
1997~\citep{fender99} and July
2001~\citep{miller05}  correspond to the beginning ($\theta\sim 0$) of
three different  period of jet precessions respectively.

Note that when the  parameter, i.e., $a_{d}$ increases,  the 3.8yr
periodicity can still hold  by assuming $\lambda>\pi/6$.

\subsection{The long-term effects}
As the bulk motion model, the jet precession model also
corresponds to different velocities of the approaching and
receding blobs. However these two models predict different
behavior on the ratios of the displacement of the approaching blob
to the receding blob. Moreover the two models also correspond to
different evolution on the flux density of the approaching and
receding blobs.

By the jet precession scenario, the speed of the blobs measured in
a frame at rest to the core is $V_0=l_0\dot{\Omega}_1\sin\lambda$.
The time measured in the observer's frame is  $T$, while in the
frame at rest to the moving blob, the corresponding time is $t$,
they are related by,
 \begin{equation}
\label{ta}
{T}=\frac{t+Vx/c^2}{(1-V^{2}/c^{2})^{1/2}}
 \ \,.
\end{equation}
Putting $V=V_{0}\sin\theta$, $x=l_{0}(1-\cos\theta)\sin\lambda$,
and $t=\theta_0/\dot{\Omega}_1$ into Eq.~($\ref{ta}$), $T_a$, the
time takes for the approaching blob moving from $\theta=0$ to
$\theta=\theta_0$ is obtained. Similarly by putting
$x=l_{0}(1+\cos\theta)\sin\lambda$ and $t=\theta_0/\dot{\Omega}_1$
into same equation, we actually obtain $T_r$, the time taken for
the receding blob rotating  by the same angle, $\theta$, as the
approaching one. Then following relation is obtained,
\begin{equation}
\label{var} T_a>t>T_r
 \ \,.
\end{equation}
By Eq.~($\ref{var}$) and  Eq.~($\ref{chobs}$), the observer on the
earth will tell that $\theta_a>\theta_r$,  and hence we have
$R_a>R_r$ by Eq.~($\ref{flux}$).

\begin{figure}[h]




\includegraphics[87, 197][700, 700]{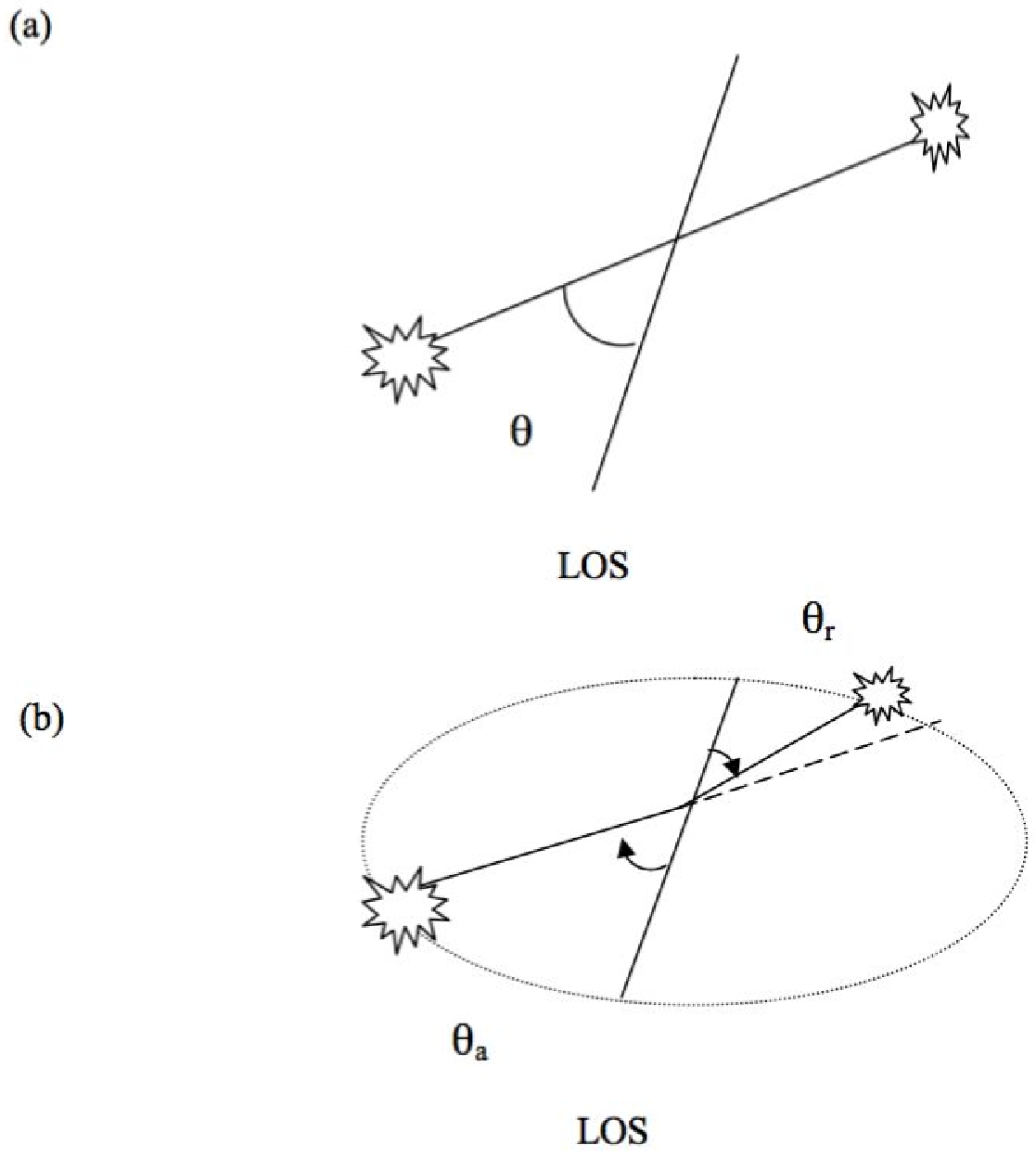}
 \caption{\label{Fig:f1} A schematic illustrating of the two models on superluminous motion
 (a) corresponds to the bulk motion model with a constant position angle
 $\theta$. Whereas,  the jet precession model corresponds to constant
 distances from the blobs to the core,  and the position angle varies
 with precession of jet axis.
 }
\end{figure}

In observer's reference frame, the time interval between the two
signals sent by the approaching blob at the moment,  $\theta=0$
and $\theta=\theta_0$, is given
\begin{equation}
\label{tta}
 \delta T_a=T_a+l_{0}(1-\cos\theta_a)\sin\lambda/c \ \,.
\end{equation}
Notice that when  $\theta=0$, the blob is at the position which
closest to the observer, as shown in  Fig.~\ref{Fig:f1}b.
Similarly time interval for the receding blob is
\begin{equation}
\label{ttr} \delta T_r=T_r+l_{0}(\cos\theta_r-1)\sin\lambda/c
 \ \,.
\end{equation}
Therefore, the apparent transverse velocities of the two blobs,
$v_a=R_a/\delta T_a$ and $v_r=R_r/\delta T_r$, can be obtained
respectively. Obviously, we have
\begin{equation}
\label{var2} v_a\sim v_r\sim V_0
 \ \,,
\end{equation}
Consequently, if $V_0\leq c$, then $v_a$ and $v_r$ should be less
than the speed of light also.

However if an observer treats the signal sent by the approaching
blob  at $\theta=0$ as from the core of the microquasar instead of
the from the position which is closest to the observer, as shown
in Fig.~\ref{Fig:f1}, then the time interval corresponding to
Eq.~($\ref{tta}$) is $\Delta
T^{\prime}_a=T_a-l_{0}(\cos\theta_a-1)\sin\lambda/c$. And the
corresponding apparent transverse velocity is
$v^{\prime}_a=R_a/\Delta T^{\prime}_a>v_a$, and hence
$v^{\prime}_a>c$ is possible. In other words, jet precession
induced blob motion may appear as superluminal motion.

Moreover,   $\theta_{a}$ and $\theta_{r}$  are different functions
of time, and by Eq.~($\ref{flux}$), $R_{r}/R_{a}$ is also a
function of time. Therefore the jet precession model predicts a
variable $R_{r}/R_{a}$.

Comparatively  the bulk motion model predict that
$R_{r}/R_{a}=1-2V_{0}\cos\theta/c$, which is a constant, since the
position angle $\theta$ and the velocity $V_0$ are constants. The
observations~\citep{mira94, fender04, miller05} likely support a
variable $R_{r}/R_{a}$.

Owing to Doppler deboosting of the receding jet  flux density,
$\theta_r=0$ corresponds to the minimum flux density. As $\theta$
increases the flux density increases,
  in the case $\theta_{r}>\theta_{cr}$ ($\theta_{cr}$
denote a critical angle),  the receding jet can be luminous enough
to be observed. This explains  the absence of X-ray emission in
the receding blob after the outburst in XTE
J1550-564~\citep{corbel02}. Whereas in the bulk motion scenario,
the absence of the receding blob is not expected, since it should
be brighter where it is near the core instead of far away from the
core.

The superluminal motion of GRS 1915+105 observed in radio band may
be dominated by the emission from the condensations instead of the
beam, therefore the effects of Doppler boosting and deboosting are
not as obvious as that of X-ray emission of XTE J1550-564.

Assuming $V_{0}=0.9c$, we have
$l_{0}=V_{0}/(\dot{\Omega}_{1}\sin\lambda)\sim 6000$mas.
Therefore, the displacement varies with  $\eta$ at a rate,
$l_{0}\dot{\Omega}_{1}\sin \lambda\sim 14$mas/day.

The major radio flare of  XTE J1550-564  in  September 1998
corresponds to the time that the eastern jet (approaching jet)
towards LOS. The observations of June, August and September 2000
correspond to the increasing  of $\theta$. The observations of
March 2002 and June 2002 show that the western (receding) blob
became much brighter than the eastern (approaching)
blob~\citep{corbel02}. This implies  $\theta>\pi/2$ in 2002, in
other words the receding blob became the approaching one.

Therefore the transition ($\theta=\pi/2$)  should happen in 2001, which
 implies a period of precession of $\sim$12yr ($\theta\sim 0$
 in September 1998, and $\theta\sim\pi/2$ in 2001). This prediction can be tested by
 further observation of  XTE J1550-564 .
 The next evolution would be the disappearance of the eastern blob and
 the displacement of the western blob  would be closer to the core
 than that in  March 2002 and June 2002. Owning to the increasing Doppler
 boosting the western blob should be brighter and brighter until the next outburst.

Two separate outbursts in 2001 March and 2001 July on GRS 1915+105
are observed~\citep{miller05}. As the former observations, no
deceleration is shown in  GRS 1915+105. Whereas in  XTE J1550-564
 deceleration has been observed~\citep{corbel02}.

 This is because most observations
of  GRS 1915+105 last a few months after an outburst, which
corresponds to a relatively small position angle variation,
$\theta\sim\pi/10$. Whereas, the  observation of  XTE J1550-564 in
2002 corresponds to  $\theta\geq\pi/2$. Therefore, XTE J1550-564
has passed the
   deceleration regime, whereas GRS 1915+105 actually has not.
  The observation of GRS 1915+105
of 1995 August 10 could have
  measured
  the return of the approaching blob, however it is unreliable since
  the individual condensations were resolved only in one epoch~\citep{mira99}.

\subsection{The short-term effects}
\label{sec:short} The precession of the jet is  actually a
short-term nutation superimposed on the long-term primary
precession. The short-term nutation corresponds to a variation of
the position angle at the time scale of the orbital period, and
the amplitude of which is  much smaller than that of the primary
precession.

By the orbital parameters we have the  orbital angular momentum,
$L\sim 4\times 10^{53}$g cm$^2$s$^{-1}$.

Assume that the mass of the ring (at radius $a_d=a/2$) is
$0.1M_{\odot}$, which corresponds to an angular momentum of
$L_d\sim 2\times 10^{52}$g cm$^2$s$^{-1}$. Thus the ratio of spin
angular momentum of the binary system to the orbital angular
momentum of the binary is $S/L\sim 5\times 10^{-2}$.  Thus the
misalignment angle, $\lambda_{LJ}\sim S/L\sim 5\times 10^{-2}$,
and hence  the orbital inclination angle $i$ of Eq.~($\ref{i}$)
and in turn $\theta$ of Eq.~($\ref{chobs}$)
can vary at the amplitude $\delta\theta\sim 5\times 10^{-2}$ rad.

Inputting the orbital parameters  (assuming $e=0$)  into
Eq.~($\ref{bo}$), the precession rate of the orbital plane  can be
obtained, $\dot{\Omega}\approx 4.4\times 10^{-6}$s$^{-1}$, which
corresponds $2\pi/\dot{\Omega}\sim 4P_{b}/5$, and at an amplitude,
$\sim 5\times 10^{-2}$ rad.

The  periodicity of $P_{b}/2$ has been found in SS433, by the spectral
lines, which is interpreted as the nodding motion of accretion rings
and disks~\citep{katz82}. This paper proposes an alternative interpretation
to the similar  periodicity,  $4P_{b}/5$ in   GRS 1915+105, which
is induced by the
rapid precession  of the orbital plane ($\dot{\Omega}$) instead of the disk.

Therefore, we have variation of  $\theta$ induced by $\Omega$ at a
period $\sim 4P_{b}/5$. And since  $\dot{\Omega}$ and  $\dot{\lambda}_{LJ}$
are function of the orbital phase, the variation of
$\theta$ also  has a period $P_{b}$.
So $\theta$ can vary at the short time scales, $\sim 4P_{b}/5$ and $\sim P_{b}$.

The observation of GRS 1915+105 in March 2001~\citep{miller05}
shows only a single northwestern component (receding), which
vibrates back and forth . The displacement from the core increases
and decreases rapidly and at large amplitude. Simultaneously three
distinct  southeastern components (approaching) were seen to be
ejected.

The three distinct approaching components can be interpreted as
multi ejections in the bulk motion scenario. Due to the Doppler
boosting, the weak components in the approaching ejectors can be
observed, whereas that of  receding ones are unobservable owing to
the Doppler deboosting. However the vibration of the single
receding component is still difficult  to understand.

By the jet precession model the three distinct approaching
ejectors might be the blob just forming plus the ones that are the
remnants of  earlier generations of bubbles.
 Owing  to the Doppler boosting of the
jet flux density, the low-temperature whorl can be observed, and
the remnant of earlier generations of bubbles might mimic
different ejections.

For the receding blob,  the  Doppler deboosting in the receding
jet flux density makes the remnants unobservable. And thus the
appearance is the  oscillation of a single receding component
(just formed).

The primary precession changes $\theta$ at a rate approximately
$\sim 0.26$deg day$^{-1}$, hence in 20-30-d, the position angle
changes $\sim 5-8$deg. And consider the contribution by the
nutation of $\sim 3$deg, the 10 deg variation on the position
angle~\citep{mira99} is possible.

If  the opening angle of the emission beam is around $10$deg, then
it would take $10/0.26=38.5$-d for the beam sweeping through LOS
by the primary precession ($\eta$). However, due to the change in
$\theta$ caused by the short-term effects, the beam can pass
through LOS before and after the primary passing process. In other
words, the actual time of sweeping  is approximately three times
of $38.5$-d, which is  $\sim 116$-d. This explains why outburst of
GRS 1915+105 lasts 3-4 months.

In one period of $\theta$, i.e., from March 1994 to October 1997,
$\theta$ may vary at an amplitude of $\theta\sim 5\times 10^{-2}$
by short-term effects.

The total change of $\theta$ by the primary precession in one
period is $\sim 1$ rad. Thus we have $\delta\theta/\theta\sim
10^{-1}$, and by Eq.~($\ref{flux}$),  the discrepancy in the
proper motion at level $10\%-20\%$~\citep{mira94, fender99} can be
explained.



 GRS 1915+105 and SS 433 show similar
 decrease in flux density with angular separation from the
core~\citep{mira99}.  By the jet precession model, after the outburst, the
position angle of  GRS 1915+105 may close that of SS433. And due to  the jets
are produced by the same physics, hence similar decrease in flux
density with angular separation is expected.

\section{The state transition of  GRS 1915+105}

The   superluminal motion of   GRS 1915+105 is the effect of jet
precession with position angle, $\theta>0$, the state transition
of it corresponds to the precession in the case  both $\theta>0$
and  $\theta\sim 0$. The precession of a structured jet may be
responsible for the complicated state transitions in GRS 1915+105.

The behavior on light-curves, color-color diagrams, and
color-intensity diagrams can be described by state A, state B, and
state C~\citep{belloni00}. States A and B corresponds to soft energy spectra
with an observable inner region of he accretion disk, whereas state C
is related to the low state~\citep{belloni97a, belloni97b}.

As in AGNs,  microquasars should also have a compact radio core
corresponding to the state, say  C1, which forms to the innermost
cone of the structured jet.

The compton scattering of accretion disk photos by relativistic
nonthermal electrons in the jet have been used to explain the
high-energy gamma radiation from extragalactic radio
sources~\citep{dermer92}.

Koerding et al (2002)
point out that if the emission is  relativistic  beamed then the
high X-ray luminosity of  an increasing number of X-ray sources in
nearby galaxies can be explained in the case of BH mass less than
10$M_{\odot}$.

Similarly the scattering of the photos from the inner region of
the disk in microquasars  may produce  beamed radiation and hence
form  cones around the compact radio core, which correspond to
high/soft  states A and B.

The inner disk may be warped like Fig.~\ref{Fig:f4}, which may be
responsible for the  discrepancy in state A and B. As shown in
Eq.~($\ref{katz}$), the outer region of the accretion disk
precesses more rapidly than that of the inner one. The precession
propagates into the inner region through viscosity, which may
contribute to the warping of the accretion disk.

On the other hand, outside the cones of states A and B is
the state C2, which is an intermediate state transforming from the beamed
to unbeamed state. State C2 is represented by the large green cone in the structured jet
as shown in  Fig.~\ref{Fig:f4}.

As the angle $\theta$ further increases  the
 beamed radiation reduces substantially.
 The emission received may be from the tilted disk,
the corona, wind from the companion star,
as well as the side of the jets, which is   much fainter
  than that of the beamed one and state C2. This low state may correspond
  to C3.
Apparently rapid variabilities can only  be produced  in the
beamed states (A and B), instead of unbeamed ones.

\begin{figure}[h]
\includegraphics[87, 87][500, 500]{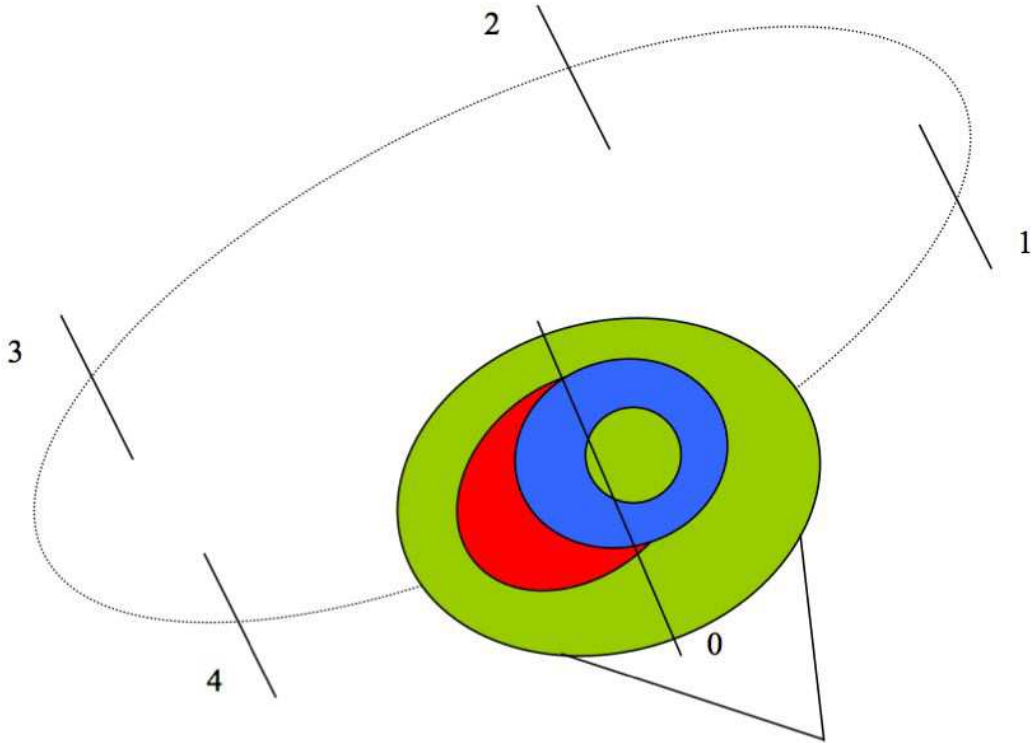}
\caption{\label{Fig:f5}
A schematic illustrating of the  jet precession in a whole period.
The label 0 corresponding to $\theta \sim 0$,  in which high state
emission is observed. States 1,  2,  3,  4 correspond to the  quiescent states.
The long-term primary precession plus the short-term nutation
may explain  the state evolution of  Fender $\&$ Belloni (2004).
}
\end{figure}

In most of time the jet axis is not aligned with LOS, so the
luminosity is low (C3), which   corresponds  to state
(i)~\citep{fender04}, as shown in Fig.~\ref{Fig:f5}. As the
precessing of the jet axis, the LOS is gradually close the jet
axis, the luminosity increases,  thus one observes state C2
(bright) corresponding to state (ii)~\citep{fender04}.

As the jet precession continues,
one would observe the state series, A, C1, A, B, and C2, as shown in
Fig.~\ref{Fig:f4}.

However due to the  short-term effect, the structured jet may
repeat the state series to the observer. This explains the
oscillation in the state transition~\citep{fender04}.

The short-term effect  in the case of low  state cannot make
significant variation in the emission, but it   changes position
angle and  vibrates the blobs  as discussed in
Section\ref{sec:short}. Therefore the oscillation in the high
state and  at the low state may be caused by the same effect of
the jet axis at different angle relative to LOS.

\section{Application to AGNs}


In the case of  a quasar the  separation  of the two blobs  is of several
million light years instead of a few light years as in microquasars.
 And the mass of black hole is of several million solar masses
 instead of  a few solar masses as in microquasars.

Assuming  $P_{b}= 10$yr, $a_{d}= 1\times 10^{15}$cm,
$m_{1}=m_{2}=1\times 10^{6}M_{\odot}$, and through
Eq.~($\ref{katz}$), $\dot{\Omega}_{1}\sim 1\times 10^{-14}$rad
s$^{-1}$ is obtained.
 In such case, we have the speed of the two blobs,
 $v\sim l_{0}\sin\lambda\dot{\Omega}_{1}\sim c$,   when $l_{0}\sim
 10^{6}$ly. If the opening angle of the beam is also 10 deg, then it would
take $\sim 6\times 10^{5}$yr for the quasar to sweep through LOS.

Therefore, the scales of length and time of the phenomena are
proportional to the mass of the BH.

Microquasars and quasars contain  three basic
ingredient, a black hole, an accretion disk heated by
viscous dissipation, and collimated jets of high energy
particles~\citep{mira98}. By an universal mechanism
for microquasars and quasars,
if the LOS is aligned with the eject matter (which can eject many times)
of  microquasars and quasars, then microblazar and blazer can be observed
respectively~\citep{mira98}.

\begin{figure}[h]
\includegraphics[87, 87][600, 600]{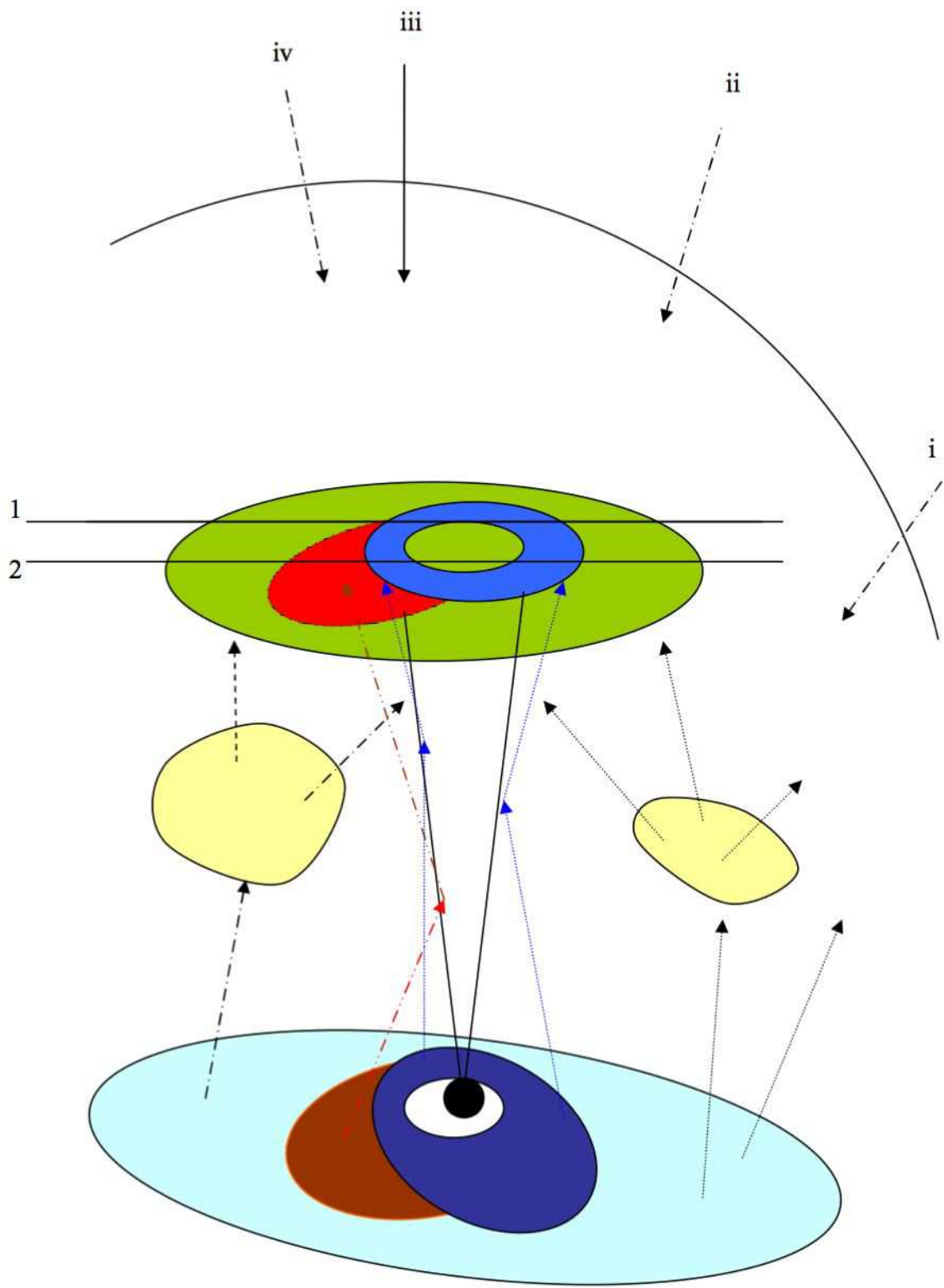}
\caption{\label{Fig:f4} A schematic illustrating of the structured
jet. The compton scattering of photos from the warped  inner
regions (blue and brown ellipse) of the disk are beamed and
responsible for the state A and B,  which can be observed in the
case iii and iv. The case $i$ corresponds to the  quiescent state,
the emission is from the disk,  the corona (the two volumes
between the disk and the beam), wind of the companion star as well
as the jet. State ii is a state between the low quiescent state
and the high active states. }
\end{figure}

 By the jet precession model, the jet is a steady out flow,
microblazar and blazar are temporary states of  microquasar and
quasar respectively.
 When the jet axis is aligned with LOS through precession,
 microblazar and blazer are observed; and when the jet axis
 precesses away from LOS, quiescent state is seen, which can
 be shown in  Fig.~\ref{Fig:f4}.

{}

\clearpage

\clearpage

\clearpage

\clearpage
\clearpage

\end{document}